# Isotope effect on the $E_{2g}$ phonon and mesoscopic phase separation near the electronic topological transition in $Mg_{1-x}Al_xB_2$

L. Simonelli[1], V. Palmisano[1], M. Fratini[1], M. Filippi[2], P. Parisiades[3], D. Lampakis[3], E. Liarokapis[3], A Bianconi[1]

[1]Department of Physics, Sapienza University of Rome, P.le Aldo Moro 2, 00185 Roma, Italy

[2]Laboratoire CRISMAT, UMR6508, 6 boulevard Maréchal Juin, 14050 CAEN Cedex4, France

[3]Department of Physics -National Technical University of Athens (NTUA), Zografou Campus, Athens GR157 80, GREECE

**Abstract**

We report the boron isotope effect on the $E_{2g}$ phonon mode by micro-Raman spectroscopy on the ternary $Mg_{1-x}Al_xB_2$ system, synthesized with pure isotopes $^{10}B$ and $^{11}B$. The isotope coefficient on the phonon frequency is near 0.5 in the full range decreasing near $x = 0$. The intraband electron-phonon (e-ph) coupling, for the electrons in the sigma band, has been extracted from the $E_{2g}$ line-width and frequency softening. Tuning the Fermi energy near the electronic topological transition (ETT), where the sigma Fermi surface changes from 2D to 3D topology the $E_{2g}$ mode, shows the known Kohn anomaly on the 2D side of the ETT and a splitting of the $E_{2g}$ phonon frequency into a hard and soft component from $x = 0$ to $x = 0.28$. The results suggest a minor role of the intraband phonon mediated pairing in the control of the high critical temperature in $Mg_{1-x}Al_xB_2$. The common physical features of diborides with the novel multigap FeAs-based superconductors and cuprates is discussed.

1. **Introduction**

High $T_c$ superconductivity, where a macroscopic quantum condensate overcomes the decoherence effects of temperature, shows up in three different systems: cuprates, diborides and iron pnictides. The physical features determining the common quantum mechanism for high $T_c$ can be unveiled by the few common features:

*first*, they are multilayer materials made of active metallic layers (boron layers in *diborides* [1-2], $CuO_2$ layers in *cuprates* [3-7] and $FeAs_{4/4}$ layers in *iron pnictides*, [8, 9]) separated by spacer layers which respectively contribute and do not contribute to the electronic states at the Fermi level;

*second,* the high $T_c$ phase shows up by fine tuning of the chemical potential in a regime where electronic states different in terms of spatial locations or symmetry coexist at the Fermi level, giving the multigap anisotropic superconductivity [7];

*third,* the high $T_c$ phase occurs in a regime of mesoscopic phase separation (MePhS) where two phases compete.

In all these systems (diborides [2], cuprates [6] and iron pnictides [9]) the chemical potential is tuned in the proximity of an electronic topological transition (ETT), where in the presence of disorder a phase separation is expected to occur [10].
The high $T_c$ superconducting diborides, where both the σ holes and π electrons coexist at the Fermi level, have been the first clear case of multiband anisotropic superconductivity [for a review see ref. 7]. In this system the duality of the electron gas at the Fermi level is provided by two 2D cylindrical Fermi surfaces of σ electrons and two 3D Fermi surfaces of π electrons. In the $MgB_2$ ($T_c \sim 40$ K) the Fermi level in the σ (π) band is close (far) to its band edge (type (I) ETT). Increasing the Fermi level a type (II) ETT is quickly reached, called opening of a neck [7, 10], where the σ surface changes dimensionality from 2D to 3D. While a strong electron phonon coupling with a Kohn anomaly (giving a phonon softening) is expected when the Fermi surface has a 2D topology [11], a suppression of this coupling is expected in the 3D regime.

In the standard BCS superconductivity mechanism phonons couple to charge carriers to give the Cooper pairs. The isotope effect on the critical temperature represents a measure to the contribution of phonons to the pairing mechanism. The isotope effect on the superconducting critical temperature (α) is expected to be directly related with the effect of isotope substitution on the frequency of the phonon mode ($α_p$). In fact $T_c$ is described by the relations $T_c \propto M^{-\alpha}$, where $M$ is the ionic mass, from the proportionality in the McMillan's formula between the $T_c$ and the mode coupled to the carriers ($\omega \propto M^{-\alpha_p}$), with $\alpha_p = 0.5$, in the harmonic approximation.

The finite boron isotope effect on the superconducting critical temperature $T_c$, measured by Bud'ko et al. [12] and Hinks et al. [13] soon after the discovery of superconductivity in $MgB_2$, has pushed the scientific community to accept the idea that this is a phonon mediated superconductor [14-16]. The very strong coupling of the B-B bond-stretching $E_{2g}$ branch of phonons to the $B$ $2p$ σ hole bonding states with 2D Fermi surface has been assumed to be responsible for the remarkable superconductivity in $MgB_2$. However the measured isotope coefficient for the critical temperature $\alpha \approx 0.3$ is less than 0.5 that is the expected value in standard BCS superconductors. Initially the reduction of the isotope effect was assigned to anharmonicity [17]. Recent experimental results and theoretical analysis indicate that anharmonicity plays only a marginal role leaving the isotope effect as the most important unresolved issue in the physics of $MgB_2$ [18].

A widely accepted idea is that the high superconducting critical temperature is mainly determined by the strength of the intraband electron-phonon pairing interaction in the σ band. In this work we provide compelling experimental evidence in contrast with this assumption. Non-conventional mechanisms have been invoked so far for high $T_c$ in diborides: bipolaron superconductivity [19], and electronic mechanisms such as the resonating valence bond mechanism [20], the electron–hole asymmetry [21], the pairing mediated by collective electronic excitations [22], charge density excitations [23, 24] and acoustic plasmons in a two component scenario [25, 26]. A two component scenario has been proposed where the key term for high $T_c$ is the exchange-like interband pairing [27-29] i.e., the direct exchange of pairs between the two components, like a "Feshbach resonance" [30-32] as discussed in ref. [7]. The resonance can be described as the direct exchange of pairs of particles in a first band and pairs of particles in a second band, where the two bands differ for spatial locations, symmetry and intraband e-ph coupling. The exchange pairing has a Feshbach resonance when the first pairs of electrons are near an ETT in the first band [7]. Therefore in diborides the high critical temperature would be controlled by the exchange-like interband pairing between the hole pairs in the σ band and the electron pairs in the π band, while the phonon mediated intraband pairing determines the pair formation. In this proposal the variation of the electron phonon coupling in the intraband pairing in the σ band is important but it is expected to have minor effects on the superconducting critical temperature.

The experimental method to test these theoretical models is to measure the response of the superconducting phase tuning the chemical potential. In fact the electron-phonon coupling is strongly sensitive to the relative position of the chemical potential and the ETT in the σ band [11].

The first method that enabled tuning the chemical potential has been the external pressure. The response of the system to pressure has provided experimental evidence of the proximity to an ETT [33]. The second method has been chemical substitutions in the spacer layers, but unfortunately substitutions in the *Mg* site appeared to be difficult, in many case unsuccessful or ambiguous. The most successful of these attempts is the *Al* substitution for *Mg*, reported by several groups [34-42], that allows the tuning of the chemical potential from above to below the edge of the σ band, while changing the superlattice misfit strain between the hcp Al/Mg layers and the boron layers that can be measured via the tensile micro-strain in the boron lattice [1].

It is now well established that the $Mg_{1-x}Al_xB_2$ ternary system is a two-band and two-gap superconductor [43-47] where the $T_c$ decreases by increasing the *Al* content, continually from 40 K in $MgB_2$ to the disappearance of superconductivity for $x$ around 0.6. The $E_{2g}$ phonon frequency shows a large softening going from $AlB_2$ to $AlMgB_4$ [48-50]. The theory of the electron-phonon interaction as function of $x$ has been developed by several authors [51-52] and the evolution of the electronic structure with $x$ has been probed by x-ray absorption and optical spectroscopy studies [53, 54].

A key point of the physics of the high $T_c$ superconductivity in diborides is that $MgB_2$, where the highest $T_c$ is reached, is at the edge of a catastrophe. In fact for small variations of the chemical potential [34] or pressure [33] the system shows a structural phase separation. For example in the case of aluminum for magnesium substitution a first structural phase separation region is detected by x-ray diffraction in the range $0 < x < 0.25$ [34,2]. However in range of doping the transport data shows a single superconducting phase with a single critical temperature indicating that the structural phase separation occurs on a smaller scale than the superconducting coherence length.

In this work we have investigated a very large number of $Al_xMg_{1-x}{}^{10}B_2$ and $Al_xMg_{1-x}{}^{11}B_2$ samples [2] using micro-Raman spectroscopy to investigate the evolution of electron-phonon interaction and the phase separation with Al doping. We have investigated the isotope pure samples as a function of the tuning of the chemical potential to detect a possible anomaly of the isotope coefficient on the phonon frequency $\alpha_p$ and the variation of lineshape of the $E_{2g}$ phonon mode in order to detect 1) the phase separation observed in high resolution x-ray diffraction but not in the superconducting critical temperature and 2) the evolution of electron-phonon interaction.

First of all here we show that the $E_{2g}$ phonon frequency follows the harmonic mass law in almost all the *Al* content range (0 < *x* < 0.57). The maximum $T_c$ is found to be at the boundary of a mesoscopic phase separation. The extracted electron-phonon coupling strength as a function of *x* shows a large increase where the chemical potential is tuned near the type (II) ETT, on the contrary it shows negligible variations where the chemical potential is tuned near the type (I) ETT in the sigma band. The results suggest that the intraband electron-phonon mechanism is not the only term controlling the high $T_c$ superconductivity in the diborides.

## 2. Experimental Methods

We have synthesized several polycrystalline samples in a wide range of *Al* content, from the pure $MgB_2$ to the ternary system with *x* = 0.57, by direct reaction method of the elemental magnesium, aluminum, and boron ($^{10}B/^{11}B$) (Eagle Picher 99% purity) [55]. The starting powders were mixed in the stoichiometric ratio and pressed into a pellet. The pellets were enclosed in a tantalum crucible, sealed by arc welding under argon atmosphere and then heated for one hour at 800°C and two hours at 950°C. The samples have been cooled to room temperature with a 4 K/m rate. Several pieces in each pellet were analyzed by XRD to look for any *Al* gradient or extrinsic inohomogeneities. We have obtained a very good reproducibility of the samples with no extrinsic inohomogeneities. The superconducting properties of all the samples were investigated by susceptibility measurements.

We have collected micro-Raman spectra on the isotopically substituted samples. The Raman spectra have been measured in the back-scattering geometry, using a T64000 Jobin-Yvon triple spectrometer with a charge-coupled device camera. The explored Raman shift range is between 50 and 1200 cm$^{-1}$. The 488.0 nm and 531.1 nm laser lines have been focused on 1–2 μm large crystallites and the power was kept below 0.1 mW to avoid heating by the beam. For each sample several measurements have been performed on different micro-crystallites choosing different region of the samples. The spectra have been collected at room temperature in a wide range of *Al* content (0 < *x* < 0.57).

3. **Results and discussion**

The micro-Raman spectra of $Mg_{1-x}Al_x{}^{11}B_2$ (filled dots) and $Mg_{1-x}Al_x{}^{10}B_2$ (open symbols) samples made of pure boron isotopes are reported in Fig. 1(a) after background subtraction. A clear phonon isotope shift is observed. To visualize the shift properly we plot the spectra of $Mg_{1-x}Al_x{}^{10}B_2$ samples as a function of Raman shift multiplied by the factor $(10/11)^{0.3}$ in panel (b) of Fig. 1 in order to compare the isotope shift of the Raman active phonon with the isotope shift of the superconducting critical temperature in pure MgB$_2$ showing an isotope coefficient 0.3 [12, 13]. In panel (c) the spectra of $Mg_{1-x}Al_x{}^{10}B_2$ samples are plot as a function of Raman shift multiplied by the factor $(10/11)^{0.5}$ according to the harmonic mass law. The spectra for the two isotopically substituted sample sets are almost coincident in Fig. 1(c), showing that the Raman response to isotopic substitution scales according to the harmonic mass law, with a slight deviation at very low *Al* content.

In the data it is possible to distinguish two phonon components: the Raman active $E_{2g}$ in-plane stretching mode of the boron atoms, and the silent $B_{1g}$ activated by disorder that involves vibrations of the boron atoms along the off plane direction. Increasing the *Al* content increase the structural disorder [2] enhancing the contribution of the $B_{1g}$ mode to the Raman spectra. At the same time, we observe a line-width narrowing and energy hardening of the $E_{2g}$-mode with increasing *Al* substitution in agreement with previous results [48-50]. In the range $0 < x < 0.28$ of *Al* content the $E_{2g}$ mode split into two $E_{2g}$ contributions (a hard mode and a soft mode) induced by the mesoscopic phase separation [2], in agreement with diffraction data showing the splitting of the *c* axis [2]. The Raman spectra have been fitted with a three component model to represent the $B_{1g}$ and the two (soft and hard) $E_{2g}$ components, as it is shown in Fig. 2. The relative weight of the soft and hard $E_{2g}$ components has been reported in Fig. 12a of Ref. 2. In Fig. 3a we report the evolution of the values of the $E_{2g}$ phonon frequency as a function of *Al* content for samples synthesized with $^{10}B$ (open symbols) and $^{11}B$ (filled dots) and their weighted average (solid, $^{11}B$, and dashed, $^{10}B$, line) of the energy of the $E_{2g}$ phonon, with the weighting given by the relative intensities of the "hard" and "soft" components. Fig. 3(b) shows the isotope shift induced on the $E_{2g}$ phonon $\alpha_p = \ln(\omega_{10}/\omega_{11})/\ln(11/10)$ hard and soft frequencies between samples synthesized with different isotopes of boron. The phonon isotope coefficient plotted as a function of *x*, shows that the average phonon isotope coefficient is $\alpha_p = 0.5$ within the error bars in the full Al substitution range. However the data for the soft mode show some possible deviations at very low doping, in fact there the error bars are much larger and the data could be consistent also with a phonon isotope coefficient as small as $\alpha_p = 0.3$ at very low Al doping.

The $E_{2g}$ phonon mode undergoes a substantial variation in terms of a weighted average, with the weighting given by the relative intensities of the "hard" and "soft" components, of the frequency (Fig. 3a) and width (see Fig. 11b in ref. [2]) below 28% of *Al* content. The investigation of a large number of samples allows us to identify the phase separation regime between 0% and 28% of *Al* content indicated by the splitting of the $E_{2g}$ mode below $x = 0.28$ in a hard and a soft mode with the relative probability plotted in Fig. 3b.

The line-width of the profiles of the soft and hard $E_{2g}$ Raman lines, in the phase separation regime, show a full width at half maximum (FWHM) around 200 cm$^{-1}$ and 150 cm$^{-1}$ respectively. So while the width decreases abruptly out of the phase separation region where the topology of the σ Fermi surface is 3D. The relative probability of the two $E_{2g}$ contributions (Fig. 12(a) in ref. [2]) shows that the weight of the soft $E_{2g}$ mode decreases on increasing the *Al* content, while the weight of the hard $E_{2g}$ mode increases. The phase separation indicated by the splitting of the Raman $E_{2g}$ mode is in agreement with the phase separation identified by the splitting of the (002) reflection peak in the x-ray diffraction data reported recently [2] confirming the early results [34]. Goncharov et al. [33] have shown that a similar phase separation is induced by non-hydrostatic pressure. They have studied the variation of the Raman spectra and x-ray diffraction (XRD) data applying either a hydrostatic or a non-hydrostatic pressure on $MgB_2$. Applying the non-hydrostatic pressure they have found a splitting of the Raman $E_{2g}$ mode, together with a splitting of the (002) XRD reflections peaks. Therefore we can deduce that the variation of the chemical pressure induced by the *Al* substitution for *Mg* induces a similar effect as non-hydrostatic pressure. The substitution of Al for Mg, in the spacer layers intercalated between the superconducting boron layers changes the misfit strain between the two types of layers forming the superlattice. Therefore we can conclude that the variation of the misfit strain has a similar effect on the splitting of phonon modes as the non-hydrostatic pressure.

In $Al_xMg_{1-x}B_2$ we observe the splitting of the phonon mode but we have not been able to detect the splitting of the superconducting transition into two different critical temperatures [55] in agreement with previous works [34,39,44,56,57] therefore we can identify the present phase separation as a mesoscopic phase separation forming a granular superconductor.

Let us now compare the present results with the case of the $Mg_{1-x}Sc_xB_2$ system showing the macroscopic phase separation in the range $0 < x < 0.1$ [58]. In this case for x>0.13 the σ Fermi surface has a 3D topology and the Kohn anomaly is suppressed as indicated by the hardening and narrowing of the $E_{2g}$ phonon mode. In this case the chemical substitution of $Sc^{3+}$ for $Mg^{2+}$ ions induces a variation of the chemical potential (the charge transfer between different layers) but since the $Mg^{2+}$ and $Sc^{3+}$ ions have the same ionic radius the misfit strain remains constant. Therefore the Sc substitution changes the charge density with a minimum lattice disorder. In the case of $Sc_xMg_{1-x}B_2$ we observe a splitting of the superconducting critical temperature indicating two different material phases with different $T_c$ [58] and therefore we can conclude that there is a macroscopic phase separation.

The phase separation occurring where the chemical potential is driven at the type (II) ETT in the boron layers is determined by the presence of two types of local perturbations in the spacers: a) the local random coulomb fields, determined the random distribution of ions with different ionic charge ($Al^{3+}$ and $Mg^{2+}$) and b) the local lattice distortions due to coexistence of ions with different ionic radius. Therefore the large miscibility gap [58] around the ETT in the case of $Sc_xMg_{1-x}B_2$ between $x = 0$ and $x = 0.15$ with the formation of a macroscopic phase separation between the $x = 0$ and the $x = 0.15$ phase is associated with the case where the role of local random coulomb fields is dominant while role of the local lattice distortions is negligible. On the contrary the mesoscopic phase separation observed here in $Al_xMg_{1-x}B_2$ is associated with the case where the local lattice distortions in the spacer layers are relevant as the random coulomb fields.

According to several works [15,31,52,53,59] the information on the strength of the intraband electron-phonon coupling (e-ph) can be extracted from the $E_{2g}$ softening [15] and line-width narrowing [11]. Since we have measured a single critical temperature in transport data we have related it with a weighted average of the energy of the $E_{2g}$ phonon, with the weighting given by the relative intensities of the "hard" and "soft" components. The weighted average of the energy (panel a ) and the FWHM (panel b) of the $E_{2g}$ phonon are plotted in Fig, 4 as a function of the *a*-axis measured by x-ray diffraction in fact the phonon energy is expected to shift with the in plane lattice compression. The expected behavior Ω(a) due to simple lattice compression for a covalent material in the absence of electron-phonon coupling is shown in Fig. 4(a) by a dashed line. The anomalous large e-ph coupling between the σ holes and the optical $E_{2g}$ phonon can be derived from the ratio between the square of the expected linear behavior Ω(a) and the square of the measured frequency that is plotted in Fig. 5(a) as a function of a-axis.

The e-ph coupling is also related with the ratio between the full width at half maximum and the energy of the $E_{2g}$ mode reported in Fig. 5(b). The frequency hardening and the line-width narrowing of the Raman $E_{2g}$ mode indicates a clear decrease of the electron–phonon intraband coupling in the σ band going from $MgB_2$ to the doped samples.

The experimental softening of the $E_{2g}$ mode is reported as a function of Al substitution in Fig. 6a in order to compare the experimental results with the predicted softening based on phonon and electronic band structure calculations of G. Profeta et al. [52], P. Zhang, [53] and O. De la Peña-Seaman et al. [59]. The results show that there is a clear disagreement between the experimental data and the theoretical predictions. The data show a relevant softening occurring only at x=0.33 (at the type II ETT, the opening of a neck) on the contrary theoretical calculations predict a relevant softening also at x=0.6 (at the type I ETT, the sigma band edge).

The intraband electron phonon coupling for the σ band holes has been extracted from the Raman data of the $E_{2g}$ mode is plotted in Fig. 6b. The electron phonon coupling has been extracted from the widening $\lambda = \Gamma/\omega/0.004 \cdot 2/9 \cdot 1/8$, according with ref. 11, and the softening of the $E_{2g}$ mode, $\lambda = ((\Omega/\omega)-1) \cdot 3/4$ according with ref. 15. In Fig.6 (b) we plot the calculated electron-phonon coupling $\lambda^* = (\lambda/(1+\lambda))$ determined from phonon softening and phonon broadening- The value of the coupling extracted from the $E_{2g}$ phonon frequency and line-width roughly agree. These results are compared with the experimental measure of the effective experimental coupling $\tilde{\lambda}_{exp} = -\ln(K_B T_c/\hbar\omega_{E_{2g}})$ obtained from the experimental $T_c$ and phonon $E_{2g}$ frequency. In fact according with the McMillan's equation $\tilde{\lambda}_{exp} = (\lambda^* - \mu^*)$ valid for $\lambda < 1.25$, where $\mu^*$ is the effective Coulomb repulsion. In the range 0<x<0.25 $\lambda^* > \tilde{\lambda}_{exp}$ so that $\mu^* > 0$ as expected. For higher Al substitution, x>0.25, where $\lambda^* < \tilde{\lambda}_{exp}$ the electron-phonon coupling cannot give account of the actual critical temperature indicating a clear role of an electronic mechanism. From these results it is clear that the measured critical temperature is considerably higher than the one calculated on the basis of the variation of the intraband e-ph coupling in the s band. Therefore the results reported here suggest that the intraband el-ph coupling seems not to be the only driving mechanism for high $T_c$ in diborides

$Mg_{1-x}Al_xB_2$ systems.

## 4. Conclusion

In conclusion we have reported the micro-Raman study of the boron isotope effect on the $E_{2g}$ phonon mode in $Mg_{1-x}Al_xB_2$ system in a wide range of *Al* content (0 < *x* < 0.57). We have found that the $E_{2g}$ phonon mode follows the normal mass law for all the *Al* contents, even if at low *Al* concentration it seems to deviate slightly from the harmonic mass law (α ~ 0.4).

In agreement with previous work [2, 33] we have detected the mesoscopic phase separation that occurs in this system as expected near an electronic topological transition of the Fermi surface. It is interesting to underline the coexistence in this system of different phases with a single superconducting $T_c$ similarly to other high $T_c$ superconductor.

Moreover we reported the measured softening and widening of the $E_{2g}$ mode, induced by the Kohn anomalies going from a 3D to a 2D Fermi surface in the isotopically pure samples. From the softening and the widening of the $E_{2g}$ mode we have calculated the electron-phonon coupling  for increasing *Al* concentration within the standard phonon mediated superconductivity, showing that it decreases much more rapidly than the experimental values. Finally we note that these results support the idea that it is possible that a common non standard BCS mechanism, involving exchange-like pairing in anisotropic multigap superconductors, is the driving mechanism for high $T_c$ superconductivity in quite different materials: cuprate perovskites [60-63], in the recently discovered iron pnictides (or FeAs multilayer) superconductors [8,9] and in the diborides. The convergence toward this unitary scenario for the pairing mechanism in high Tc superconductors is based on the common features [64-70] appearing in quite different systems. We have shown here that diborides share, with other high $T_c$ superconductors, the typical feature of being in a region near the edge of phase separation. It is possible that this is at the quantum critical point of a first order phase transition [3,9,68] driven by two variables, first, the electronic doping in the active planes and, second, the elastic misfit strain between the planes in the multilayer.

**Acknowledgments:**

Thanks are due to Prof. Robert Markiewicz for very useful discussions. We acknowledge financial support from European STREP project 517039 "Controlling Mesoscopic Phase Separation" (COMEPHS).

*Figure captions*

Fig. 1: Comparison between Raman spectra on $Mg_{1-x}Al_x^{11}B_2$ (filled dots) and $Mg_{1-x}Al_x^{10}B_2$ (open symbols) samples for $0 < x < 0.57$ (panel a). The energy scale of the Raman spectra of $Mg_{1-x}Al_x^{10}B_2$ (filled dots) is multiplied by the factor $(10/11)^{0.3}$ (panel b) and by the factor $(10/11)^{0.5}$ (panel c).

Fig. 2: Typical fit with three components of the Raman data of $Mg_{1-x}Al_x^{11}B_2$ (filled dots) and $Mg_{1-x}Al_x^{10}B_2$ (open symbols) samples (the Raman shift of $Mg_{1-x}Al_x^{10}B_2$ is multiplied by the factor $\sqrt{10/11}$). The dashed lines represent the $B_{1g}$ contribution, the solid line the soft $E_{2g}$ contribution and the dotted line the hard $E_{2g}$ contribution for aluminum concentration 20%, 10% and 6% showing the increasing intensity of the hard $E_{2g}$ component with increasing Al concentration.

Fig. 3: Panel a). The frequency of the hard and soft components of the $E_{2g}$ mode for $^{10}B$ (open circles) and for $^{11}B$ (filled dots) is reported. The energy hardening of the $E_{2g}$-mode with increasing Al substitution is shown. The splitting of the hard and soft $E_{2g}$ mode (phase separation) occurs between 0% and 25% of Al content. The weighted average of the $E_{2g}$ frequency weighted by relative intensity of the "hard" and "soft" components (reported in Fig. 12a of Ref. 2) are plotted for the $^{11}B$ systems (solid line) and $^{10}B$ systems (dashed line). Panel b). The isotope coefficient of the phonon mode as a function of $x$. The open symbols correspond to the low energy (soft) mode, while the filled dots to the high energy (hard) mode.

Fig. 4: (a) Energy of the $E_{2g}$ mode as a function of the $a$-axis going from the $AlB_2$ to $MgB_2$ samples. The dashed line shows the expected behavior due to lattice expansion for a metallic covalent material. The open circles represent the mean energy of the $E_{2g}$ mode for the $Al$ doped system. (b) The line-width of the $E_{2g}$ phonon mode as a function of the $a$-axis. The open circles represent the mean value of the $E_{2g}$ line-width for the $Al$ doped system.

Fig. 5: (a) Ratio between the expected frequency due to the variation of the lattice compression and the measured $E_{2g}$ phonon frequency as a function of the $a$-axis in the $Mg_{1-x}Al_xB_2$ system. (b) Ratio between the line-width and the energy of the $E_{2g}$ phonon mode as a function of $a$-axis. The black squares correspond to the soft $E_{2g}$ contribution, while the black disks to the hard $E_{2g}$ contribution. The calculated mean values are reported in open circles.

Fig. 6: (a) The measured $E_{2g}$ phonon frequency $\omega_{E_{2g}}$ normalized for the frequency shift due to lattice $a$-axis compression $\Omega_{E_{2g}}$ in the $Mg_{1-x}Al_xB_2$ system (filled dots). The experimental softening of the $E_{2g}$ mode frequency as a function of Al substitution is compared with calculations of Profeta et al. in ref. 52, Zhang et al. in ref. 53 and De la Peña-Seaman et al. in ref. 59. (b) The electron-phonon coupling $\lambda^* = (\lambda/(1+\lambda)$

calculated from the softening [15] (open circles) and widening [11] (open squares) of the $E_{2g}$ mode. The experimental effective coupling $-1/\ln(K_B T_c/\hbar\omega_{E_{2g}})$ (filled circles) obtained from the ratio between the superconducting critical temperature $T_c$ measured by susceptibility measurements [55] and the average $E_{2g}$ phonon frequency.

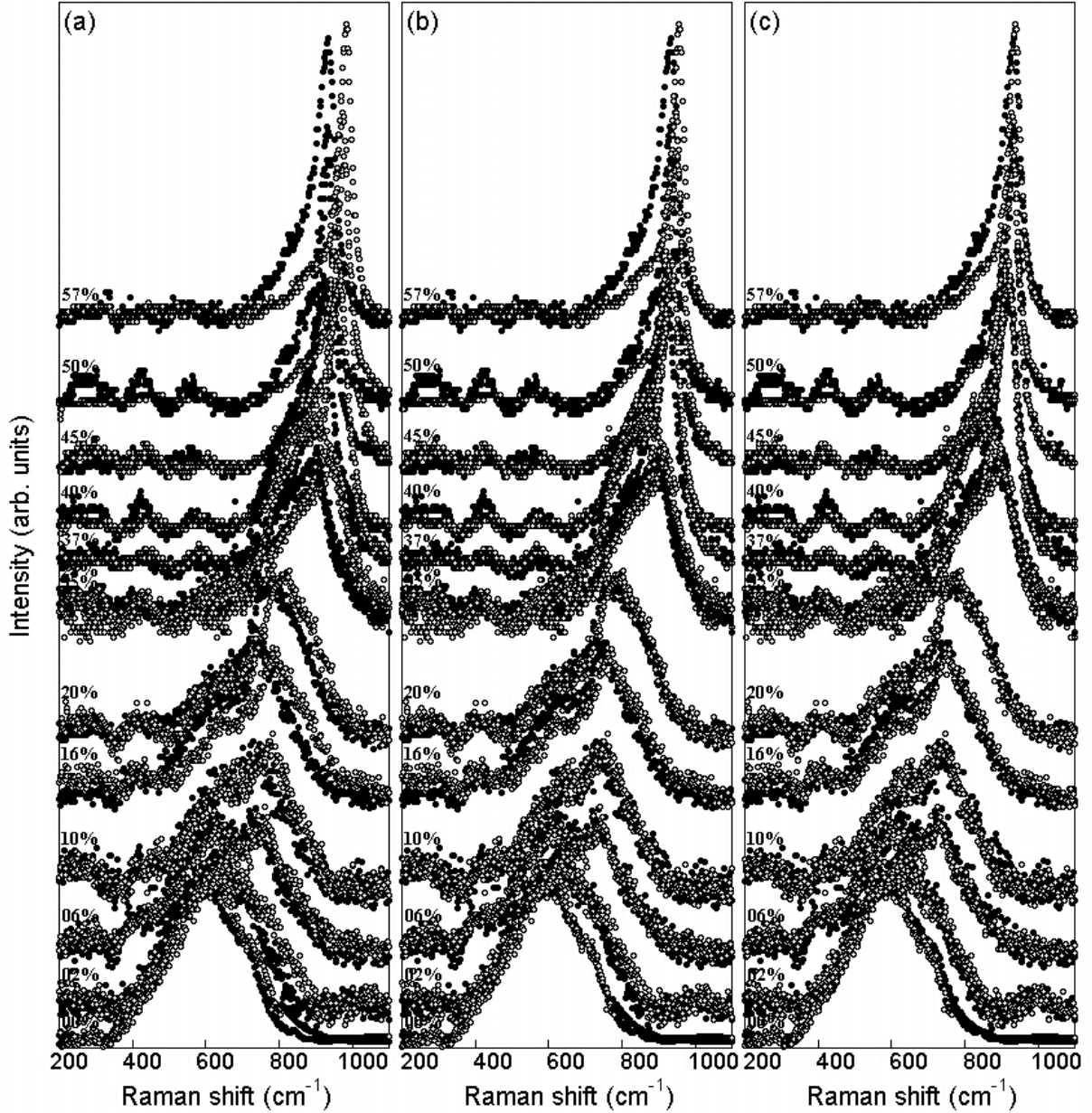

Fig. 1

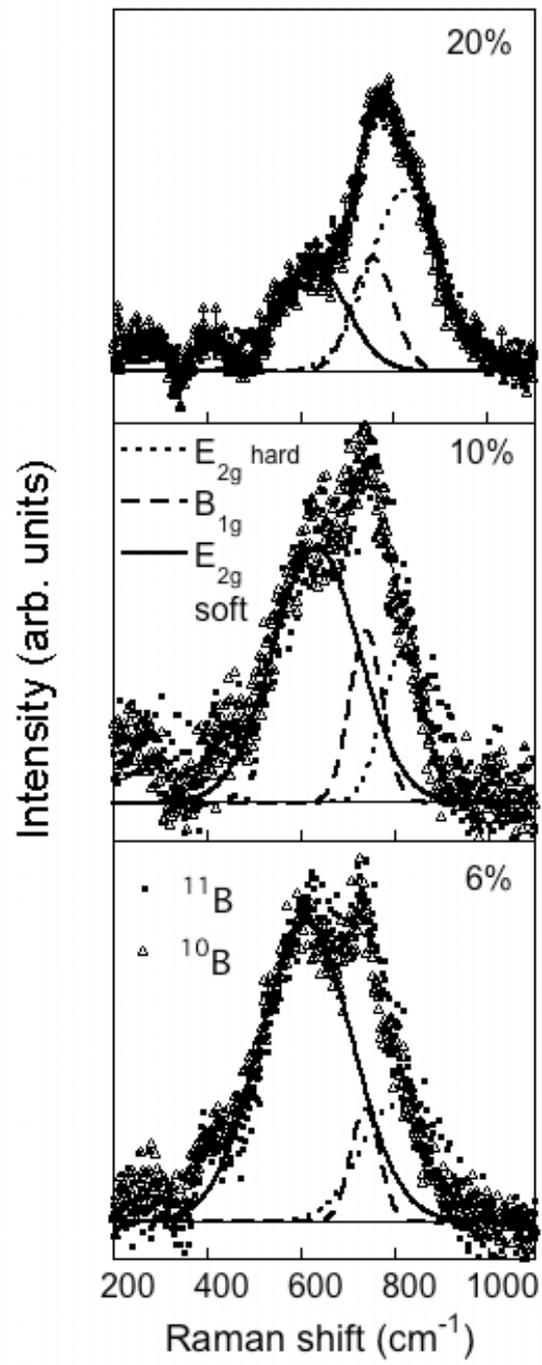

Fig. 2

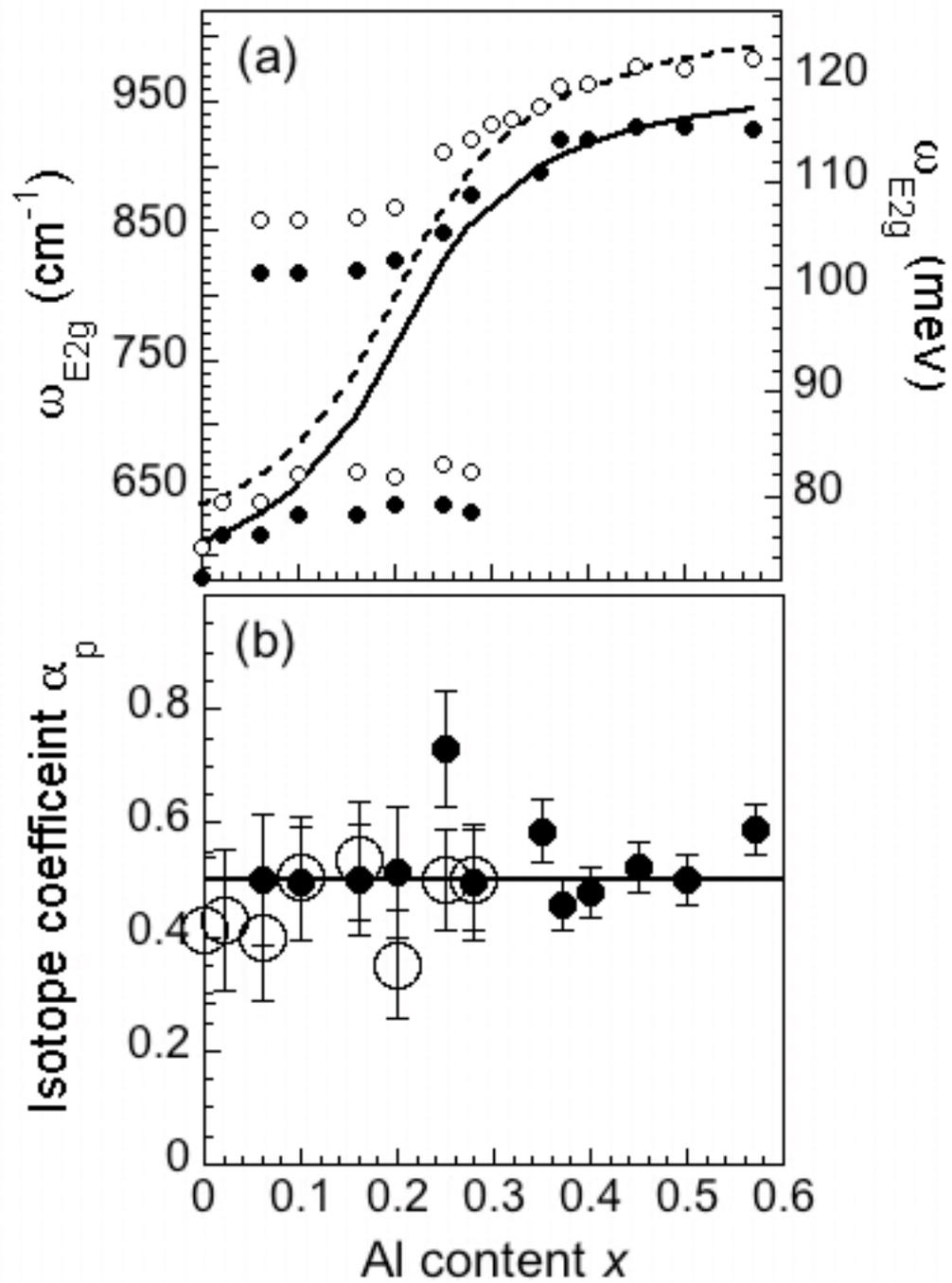

Fig. 3

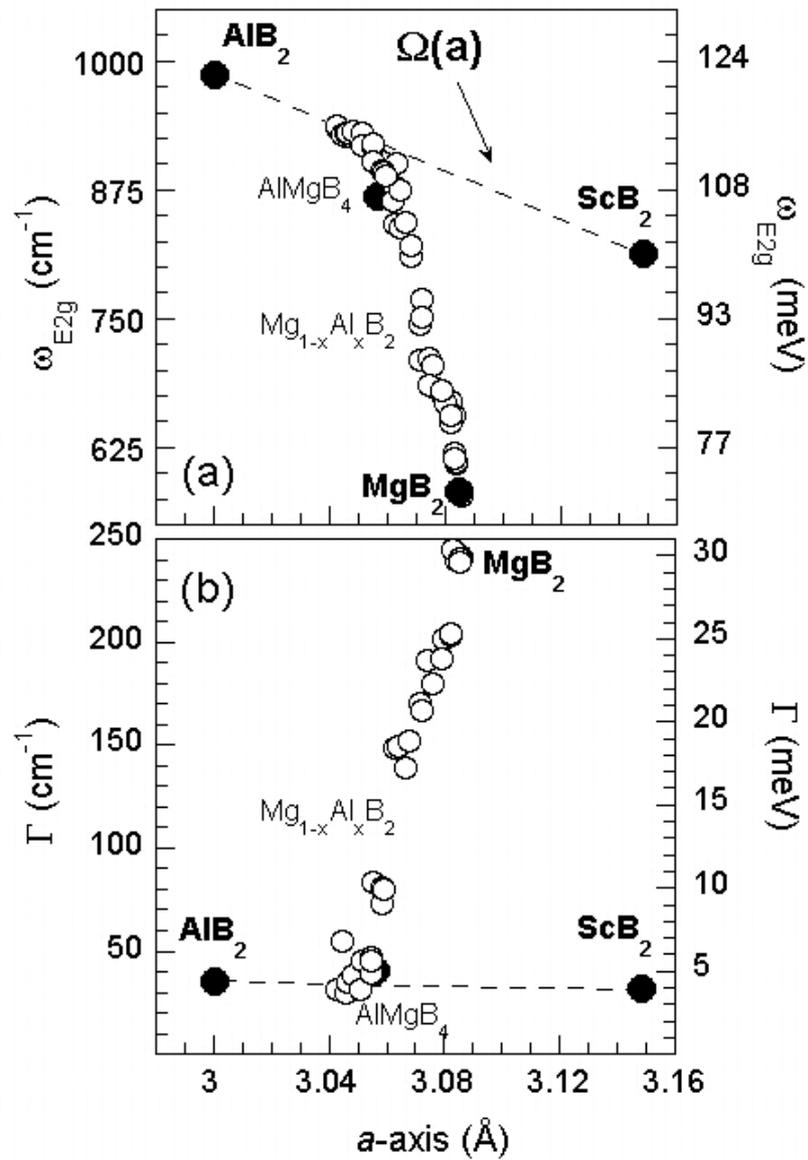

Fig. 4

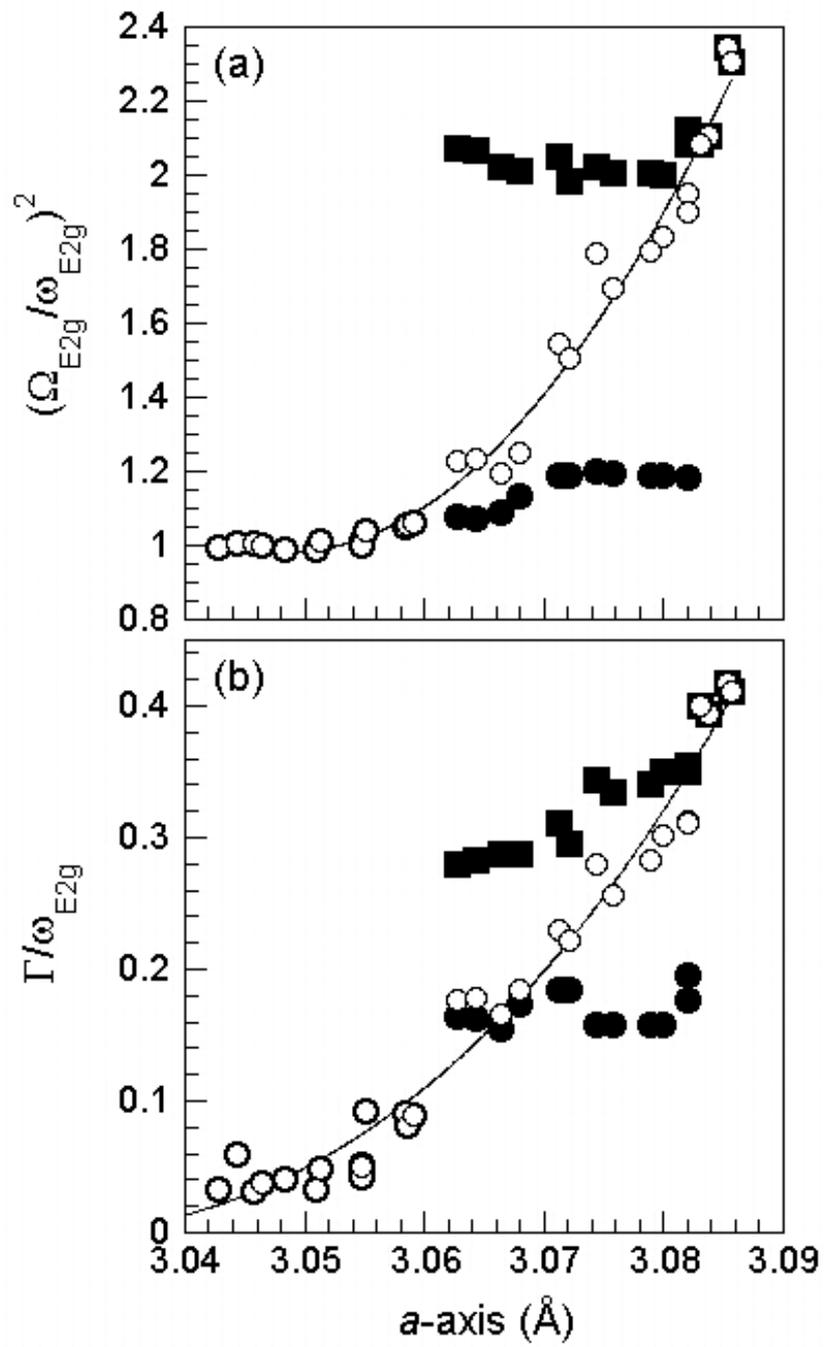

Fig. 5

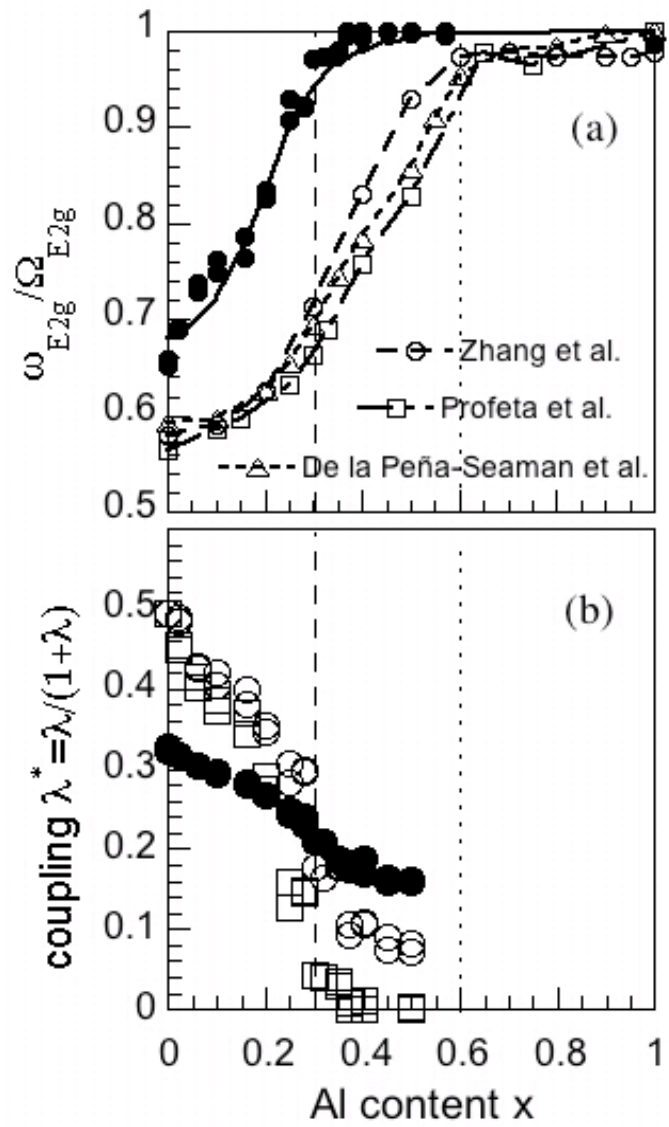

Fig. 6